\documentclass{article}

\usepackage[numbers]{natbib}

\usepackage[final]{neurips_2021}

\usepackage[utf8]{inputenc} % allow utf-8 input
\usepackage[T1]{fontenc}    % use 8-bit T1 fonts
\usepackage{hyperref}       % hyperlinks
\usepackage{url}            % simple URL typesetting
\usepackage{booktabs}       % professional-quality tables
\usepackage{amsfonts}       % blackboard math symbols
\usepackage{nicefrac}       % compact symbols for 1/2, etc.
\usepackage{microtype}      % microtypography
\usepackage{xcolor}         % colors
\usepackage{graphicx}
\newcommand{\code}[1]{{\texttt{#1}}}

\title{DAG Card is the new Model Card}

\author{%
  Jacopo Tagliabue\thanks{Jacopo designed the first card prototype, which kick-started the project; Ville created Metaflow while working at Netflix and he is one of its main maintainer; Ciro worked on the paper draft and researched the context for the article; finally, Valay implemented cards in Metaflow.} \\
  Coveo Labs\\
  \texttt{jtagliabue@coveo.com} \\
  \And
  Ville Tuulos \\
  Outerbounds\\
  \texttt{ville@outerbounds.co} \\
  \AND
  Ciro Greco \\
  Coveo Labs\\
  \texttt{cgreco@coveo.com} \\
  \And
  Valay Dave \\
  Outerbounds\\
  \texttt{valay@outerbounds.co} \\
}

\begin{document}

\maketitle

\begin{abstract}
  
 With the progressive commoditization of modeling capabilities, data-centric AI recognizes that what happens before and after training becomes crucial for real-world deployments. Following the intuition behind \textit{Model Cards}, we propose \textit{DAG Cards} as a form of documentation encompassing the tenets of a data-centric point of view. We argue that Machine Learning pipelines (rather than models) are the most appropriate level of documentation for many practical use cases, and we share with the community an open implementation to generate cards from code.

\end{abstract}

\section{Introduction}
\label{intro}

While software is eating the world, the growing use of machine learning (ML) in production systems makes it increasingly hard to understand exactly~\textit{how}. Even setting aside interpretability issues \cite{10.1145/3236386.3241340}, ML models are harder to test and debug compared to traditional software. To this extent, \citet{10.1145/3287560.3287596} introduced \textit{Model Cards}, i.e. short documents accompanying models that provide information on accuracy, biases, limitations and best practices for their use. Model Cards are meant to provide a ``reference for all, regardless of expertise'' \cite{FaceCard}: ``ML folks'' may check details on architecture and metrics, ``product folks'' get to know strengths and weaknesses of the model in different scenarios, while final users -- when applicable -- can understand better how model decisions impact their lives \cite{Seifert2019TowardsGC}.

Recently, proponents of data-centric AI \cite{NG,ReGithub} raised awareness about the importance of what surrounds model choices as well, arguing that what happens before and after training is as important as modelling \cite{DBLP:journals/corr/abs-2001-08361}. Take the latter point at face value, we believe that documentation should incorporate those aspects as well. In \textit{this} work, we motivate the move from Model Cards to \textit{Directed Acyclic Graph (DAG) Cards}, and share with the community an implementation for \textit{Metaflow} \cite{Metaflow}, the popular framework for ML workflows. 

\section{From models to pipelines}
\label{pipelines}
Data-centric AI sits at the conjunction of two theses: on one side, modelling has been increasingly commoditized \cite{10.1145/3475965.3479315}, as deep learning's inherent flexibility and pre-trained models provide good out-of-the-box performances; on the other, the Cambrian explosion \cite{Chip} of MLOps tools brought more attention on the fundamental importance of what happens before (data collection, preparation, labeling, quality \cite{Rogers2021ChangingTW}) and after (testing, serving, monitoring, drifting) the modelling stage \cite{49953,Tagliabue2021YouDN}.

\begin{figure}[ht]
  \centering
  \includegraphics[width=10cm]{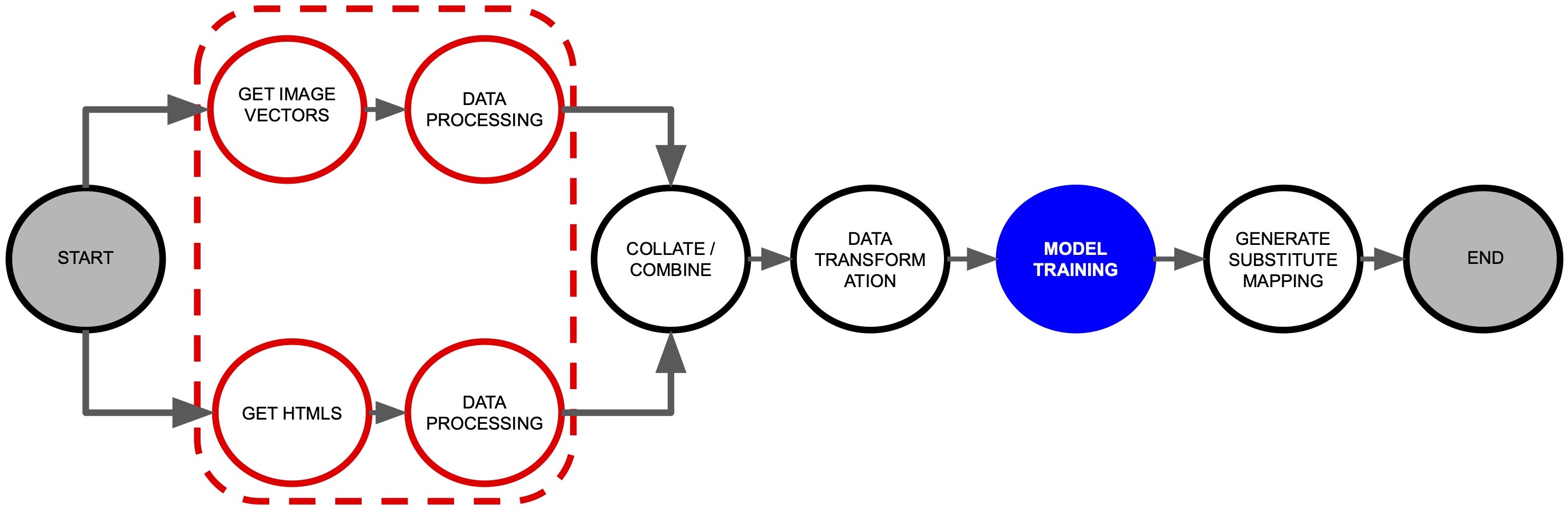} 
  \caption{DAG for the e-commerce substitute model proposed by \citet{areUSure}. Tasks in \textit{red} are run in parallel, tasks in \textit{blue} are selectively executed on GPU-accelerated hardware.}
  \label{fig_pipeline}
\end{figure}

While the original cards~\cite{FaceCard} showcased a B2C scenario, where APIs are public and use cases require no specific knowledge, Fig.~\ref{fig_pipeline} shows a more typical setting for ML developers: we borrow the pipeline from \citet{areUSure} -- a recent research paper built with Metaflow -- to have a specific example in mind, but the structure is fairly common to many modern pipelines \cite{TagliabueGithub}. As any practitioner knows, ``training a model'' involves stitching together a heterogeneous set of functionalities, from gathering data to cleaning / aggregating it, from preparing features to evaluating performances. In other words, ``training a model'' is not an atomic operation, but a series of tasks with explicit dependencies: some tasks need to be executed before others, some can run in parallel; failures may require restarting some tasks, but not all of them -- and so on. 

Conceptually, we move from the typical research setting of stand-alone scripts, to the more structured concept of DAG: while dependency management was present in earlier frameworks as well \cite{LuigiGithub}, Metaflow goes one step further, and provides the possibility of picking \textit{per task} a particular combination of Python packages and computational resources. 

Data-centric AI understands that all tasks -- not just modelling -- play a crucial role in determining the final behavior of the system, when deployed: if complex software systems have been long evaluated through ``behavioral testing'' \cite{536464}, \citet{Ribeiro2020BeyondAB} argue that it is time for ML to do the same \cite{recListPre2021}. In other words, given that data quality, feature preparation and choice of metrics influence the output of the system in the wild, properly documenting models involves documenting the entire ML pipeline. 

\section{Modern ML pipelines with Metaflow}
\label{metaflow}

Metaflow was originally developed at Netflix and it is now available as an open source package at \url{https://github.com/Netflix/metaflow}. By providing a Python-first, local-to-cloud integrated environment, Metaflow nudges users into following data-centric best practices through its design. In particular, Metaflow provides three key features to support reproducible and robust ML pipelines: %It is designed to easily allow ML developers to develop, test, and deploy an end-to-end ML workflow in a systematic manner, so that the resulting application meets requirements of production-grade ML. 
%We recognize that production-grade ML needs to address two separate sets of concerns: first, concerns that apply to any modern software deployments in general, such as high-availability and scalability, and second, the operational and philosophical requirements of data-centric AI.

\begin{enumerate}
    \item \textbf{workflow definition through DAG}: by structuring the application as a DAG-based workflow, it replaces sets of \textit{ad hoc} scripts and unreliable piecemeal execution. Furthermore, Metaflow introduces a shared lexicon for the team to model and discuss pipelines in a consistent way: \emph{Flows} (the DAG), \emph{Steps} (nodes of the graph), \emph{Tasks} (units of execution yielded by steps), and \emph{Data artifacts} (the task state). When a flow is executed either manually or by a scheduler, the execution is tracked as a \emph{Run}.
    \item \textbf{Reproducible, isolated execution of each step}: Metaflow packages together user-defined code and 3rd party dependencies in an immutable code package, which can be shipped and executed on various compute layers. Variables (such as data and states) are also automatically stored (see below).
    \item \textbf{Abstraction of computation and scheduling}: Metaflow comes with built-in integrations to popular cloud-based batch compute layers, such as AWS Batch or Kubernetes, and schedulers, like AWS Step Functions. %This allows ML developers to seamlessly switch between local and remote computing and, for example, delegate to cloud GPUs only the \textit{steps} involving training a deep neural network.
\end{enumerate}

Moreover, Metaflow and data-centric AI share the core assumption that the final behavior of ML systems is due to the interaction of two equally important components: modelling \textit{and} data. As such, Metaflow ships with additional features that make it an ideal tool to support automated generation of DAG Cards: 

\begin{enumerate}
    %\item Built-in primitives for \textbf{data flow and state management}. ML workflows are fundamentally a combination of data and compute, and hence it makes sense to treat the data flowing through the DAG as a first-class citizen.
    \item \textbf{High-throughput access to large datasets} - ML workflows are data-intensive by nature and it is common for functions to be IO-bound. Metaflow provides tools and patterns for quickly moving large datasets from a cloud-storage directly to the process memory.
    \item \textbf{Snapshotting and versioning of all artifacts}: code, results and execution metadata are automatically tracked and persisted in an immutable data store \cite{43864}; dataset, features, model weights can be  stored and versioned as well using built-in functionalities. While the importance of version control for the software/modelling aspect of a pipeline was recognized a long time ago, Metaflow emphasizes recording and tracking all inputs, intermediate states and data features as well, for full reproducibility.
    \item \textbf{Native support for experimentation}: if it is true that data-centric AI is a fundamentally iterative activity, %-- and therefore that ``production-ready'' is indeed \textit{a spectrum}, rather than a boolean flag --, 
    it is crucial to prototype quickly, without sacrificing reproducibility and rigor. Metaflow supports artifact inspection and \textit{ad hoc} analysis through notebooks, and it can accommodate the use of any off-the-shelf library for data analysis, data QA and modelling, ensuring that data scientists are able to choose the best approach for each use case.
    %that are extremely helpful for data exploration and analysis of results. Notebooks by themselves lack many features required by production-grade ML as listed above, but they allow quick adhoc analysis of results, providing a layer of observability for production workflows.
   % \item \textbf{Support for rapid prototyping} that is a key requirement of empirical, . An important element of prototyping is the ability to use any off-the-shelf libraries in the workflow, ensuring that the data scientist is able to choose the best modeling approach for each use case.
\end{enumerate}

\section{DAG Cards}
\label{dag}

\begin{figure}[ht]
  \centering
  \includegraphics[width=14cm]{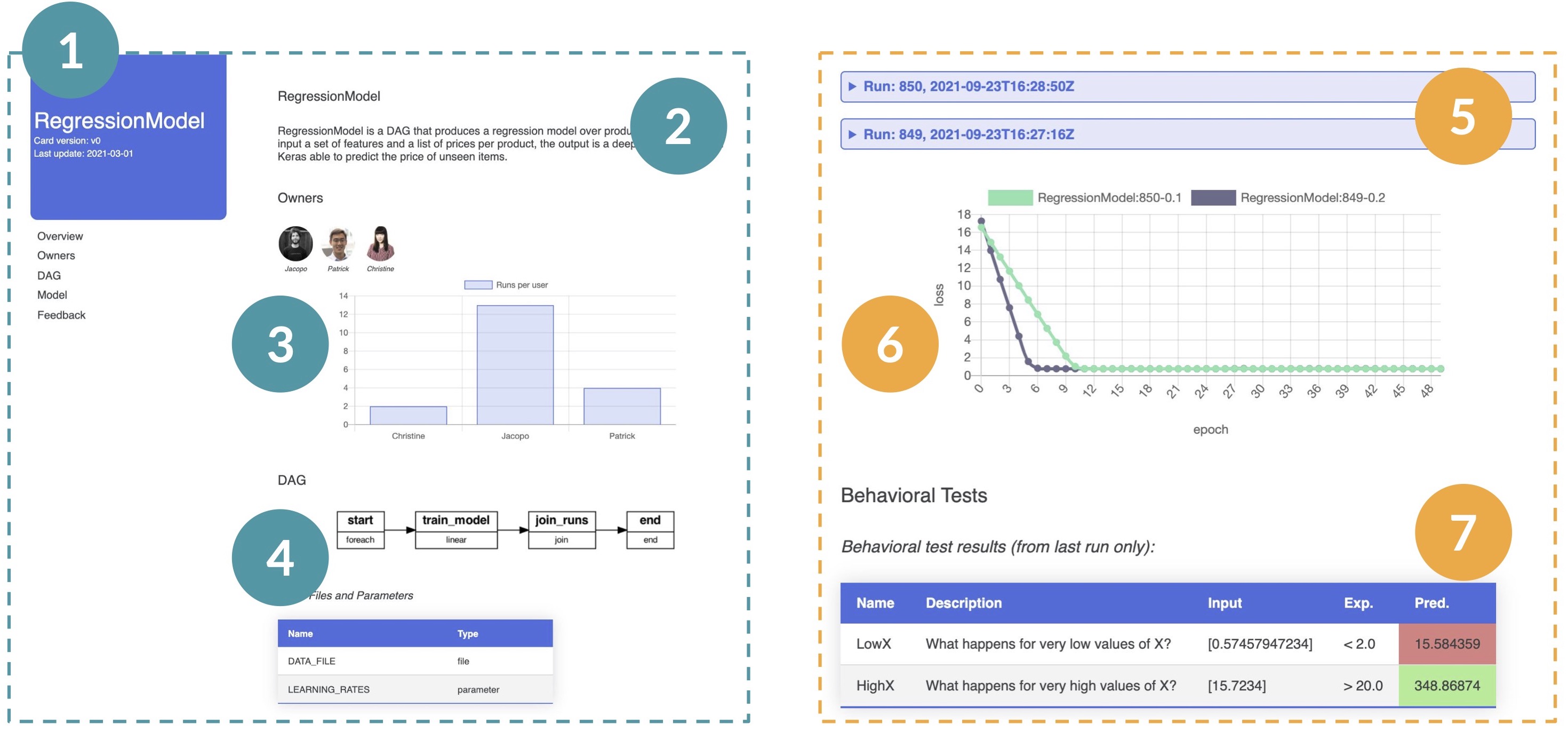} 
  \caption{Sample of DAG Card rendering from the code by \citet{cardGithub}: on the left, Flow-level information (green); on the right, Run-level data (orange). Note that all details are automatically generated by comments, code introspection and third-party APIs, with no manual intervention.}
  \label{fig_card}
\end{figure}

We designed \textit{DAG Cards} as a team of practitioners leveraging our extensive combined experience in ML for e-commerce, customer service and the entertainment industry. It is easy to realize the advantages of DAG Cards, especially when compared to popular tools of knowledge management that are ubiquitous in the industry: on the one hand, traditional software documentation - such as API docs -, on the other, internal product memos - such as intranet wiki-like pages. Based on our experience transitioning from these approaches to cards, we wish to highlight the following usability and philosophical principles:

\begin{enumerate}
    \item \textbf{low effort}: cards require a small effort from the developer -- as long as the code is well-commented/decorated, changing the code automatically results in new, up-to-date documentation. Embracing the ``documentation-as-code'' idea has two intended consequences: on the organizational side, we don't promote the hand-over of explaining a system from engineers to PMs, but instead require ML developers to be responsible for (at least some of) the explanatory artifacts; on the practical side, we remove the additional step -- often neglected, postponed, forgotten -- of updating some \textit{other} website after pushing working code; every time this step is forgotten, documentation gets stale, less useful and in some cases even potentially harmful;
    \item \textbf{versioning}: cards are artifacts themselves and can be generated after each run with pointers to the dataset, model weights, and all other artifacts in use; every card can be stored as an immutable, self-sufficient record of a particular run; as such, practitioners are able to quickly know how to debug specific states or even re-run the Flow entirely;
    \item \textbf{extensibility}: cards can be extended to incorporate all sorts of information (including interactive elements) and, more generally, act as a one-stop repository for data stored in multiple tools (e.g. retrieve accuracy from an experiment tracking system). Not only internal wiki pages are static, but they often paint an incomplete picture, as it is not obvious that authors have access to all the third-party tools storing the required contextual data.
\end{enumerate}

To make things more concrete, Fig.~\ref{fig_card} displays some selected features of a sample DAG card, as a vanilla web page so it would be easy to store and to share through a web browser. Flow-level data are in \textit{green}, Run-level data in \textit{orange}:

\begin{enumerate}
    \item \textbf{Title and menu}: the name of the Flow implicitly names the card and provides quick access to the card sections;
    \item \textbf{DAG description}: the section describes what the pipeline is for and provides contextual information; it is rendered automatically through Python introspection, i.e. using \code{obj.\_doc\_} to retrieve the class docstring.
    \item \textbf{Ownership}: the section provides information about the users working on the DAG and plots the distribution of the runs to get a sense of the relative involvement of the team members.
    \item \textbf{Structure and parameters}: the section prints out the DAG for easy visual inspections of the dependencies and tracks the parameters and input files determining the pipeline behavior; files are automatically tracked by Metaflow and can easily be linked to the DAG card for speedy lookup.
    \item \textbf{Model architecture and training info}: the section reports for the last \textit{k} (here $k=2$) runs specific info about training (e.g. training and validation loss, as recorded by a model-independent third-party tool \cite{wandb}) and a serialized representation of the chosen architecture (e.g. the output of Keras \code{model.summary}).
    \item \textbf{Loss (or accuracy) per epoch chart}: using data from a third-party tool, the section charts the loss/epoch chart for the last \textit{k} runs.
    \item \textbf{Behavioral tests}: ML systems should be stress-tested on input-output pairs of particular interest in a ``black box'' fashion \cite{recListPre2021,Ribeiro2020BeyondAB}. Behavioral tests can, for example, highlight performance deficits on a subset of users (in analogy with CATE in A/B testing \cite{Wong2019EfficientCO}), check for regression errors, control for edge cases and act as sanity checks against known problematic inputs. Of all the sections, this is the one encouraging collaboration between engineering and product teams the most.
\end{enumerate}

Following a stand-alone prototype \cite{cardGithub}, we are now releasing a first implementation of DAG Cards within the official \textit{Metaflow} codebase. The same approach we champion can obviously be extended to other DAGs: by open-sourcing our implementation of the above principles, we hope to provide practical guidance to help extending cards to other tools in the community.

\section{Conclusion}
\label{conclusion}

We argued that ML pipelines are the most appropriate level of analysis for ML systems, and surveyed what makes Metaflow suitable for data-centric AI workloads. Leveraging its abstractions, we propose a first open-source implementation of DAG Cards, and highlight important principles so that the same features could easily be ported to other tools. 

If, as argued in \citet{10.1145/3475965.3479315}, the next wave of ML systems will truly allow a substantially larger number of people to train and deploy models, documenting \textit{pipelines}, not just (commoditized) models, will be more and more important for all the stakeholders involved. DAG Cards are just a first step towards this future.

\begin{ack}
Authors wish to thank two anonymous reviewers, Andrea Polonioli, Luca Bigon and Patrick John Chia from Coveo, as well as Parker Barnes from Google, for precious feedback on earlier versions of this work.
\end{ack}

\bibliographystyle{abbrvnat}
\bibliography{refs}

%%%%%%%%%%%%%%%%%%%%%%%%%%%%%%%%%%%%%%%%%%%%%%%%%%%%%%%%%%%%

%% we don't need it for the workshop, so commenting it out
%%\section*{Checklist}

\end{document}